\documentclass[conference]{IEEEtran}
\IEEEoverridecommandlockouts
\usepackage{cite}
\usepackage{amsmath,amssymb,amsfonts}
\usepackage{algorithmic}
\usepackage{graphicx}
\usepackage{textcomp}
\usepackage{array}
\usepackage{hyperref}
\usepackage{subfig}
\usepackage{xcolor}
\usepackage{hyperref}
\hypersetup{
    colorlinks=true,
    citecolor=blue,
    linkcolor=blue,
    filecolor=magenta,      
    urlcolor=cyan,
    }
\renewcommand\thesectiondis{\arabic{section}}
\renewcommand\thesubsectiondis{\thesectiondis.\arabic{subsection}}
\renewcommand\thesubsubsectiondis{\thesubsectiondis.\arabic{subsubsection}}

\def\BibTeX{{\rm B\kern-.05em{\sc i\kern-.025em b}\kern-.08em
    T\kern-.1667em\lower.7ex\hbox{E}\kern-.125emX}}
\begin{document}

\title{A Systematic Literature Review on Smart Contracts Security\\
}
\author{\IEEEauthorblockN{ Harry Virani }
\IEEEauthorblockA{\textit{Department of Engineering} \\
\textit{University of Guelph}\\
Guelph, Canada \\
hvirani@uoguelph.ca}
\and
\IEEEauthorblockN{ Manthan Kyada }
\IEEEauthorblockA{\textit{Department of Engineering} \\
\textit{University of Guelph}\\
Guelph, Canada \\
mkyada@uoguelph.ca}
}
\maketitle

\begin{abstract}
Smart contracts are blockchain-based algorithms that execute when specific criteria are satisfied. They are often used to automate the implementation of an agreement so that all parties may be confident of the conclusion right away, without the need for an intermediary or additional delay. They can also automate a process so that the following action is executed when circumstances are satisfied. This study seeks to pinpoint the most significant weaknesses in smart contracts from the viewpoints of their internal workings and software security flaws. These are then addressed using various techniques and tools used across the industry. Additionally, we looked into the limitations of the tools or analytical techniques about the found security flaws in the smart contracts.
\end{abstract}


\section{Introduction}
With the use of distributed ledger technology (DLT), individuals with little to no confidence in one another may trade any kind of digitized information peer-to-peer (P2P) using few to no middlemen \cite{r6}. In this sense, it replaces conventional middlemen or reliable third parties, at a minimum. The certain transaction or asset that may be transformed into electronic form, such as currency transactions or storage, health records, birth, marriage, and insurance certificates, the purchase and sale of products and services, and insurance contracts, could be represented by the data transferred.

A subclass of DLTs called block-chain uses "blocks" of data to record data transactions over a distributed network of many nodes or computers. Party A asks for a transaction with party B, such as a money transfer, a contract, or the exchanging of documents. This transaction is sent out to a dispersed network of "nodes," or computers, who will validate it in accordance with a set of predetermined guidelines known as a "consensus" method. An additional "block" will be added to the block-chain once the transaction has been verified. 16 A pointer to the previous block in the chain is supplied, the transaction data is submitted, and the new block is timestamped when it is added to the block-chain.

Then, the cryptographic technology is used to process data where a hash is produced based on the hash of the fresh block's data contents plus those from the preceding block. The final result then becomes the new block's hash. Through this procedure, each block is connected to the one before it, making a chain of blocks (thus the the term "block-chain"). Each node or computer in the network contributes a unique record to the block-chain.
and is always synchronised and updated. Block-chain finally maintains the records as a database or ledger of every transaction carried out across the network.

\subsection{Prior Research}
According to the article given in \cite{r1}, their study is focused on the Document of Understanding (DOU) contract, which is the foundation of the partnership between a consumer service and its supplier. It is directed at supply chain activities. There is a chance to use blockchain technology as a solution because the approval process for supply chain activities is currently taking too long \cite{a1}. They utilised regional resources for our project. Creating a localised blockchain ledger using resources, agile methods, and design thinking. As a consequence, they created a proof-of-concept Blockchain prototype that promotes secrecy and preserves participant private information while having the whole history of the agreement, including immutable transactions. With this demonstration, they measured the time required to obtain the DOU contract's approval from all parties involved, and it was significantly reduced. The project's original contribution is implementing Blockchain in our company's operations, which enhances business processes and provides staff with a real-time view of all the data. As a consequence, their business operations have significantly improved when they combine their work processes with cutting-edge technology. Now that the program has had a successful test run, they can confidently implement smart contracts in regular Smart City operations. It may also be used to other professions that deal with financial reporting and private data.

Apart from that paper,  we also found a study \cite{r2}, they provide an automated deep learning strategy to learn the structural code embeddings of smart contracts in Solidity, which is important for contract validation, clone identification, and bug detection on smart contracts.they apply our methodology to more than 22K Solidity contracts obtained from the Ethereum blockchain, and the findings demonstrate that Solidity code has a substantially higher clone ratio (about 90\%) than conventional software. As our bug database, we compile a list of 52 recognised flawed smart contracts that fall under 10 categories of widespread vulnerabilities. Using our bug databases, the method can effectively and precisely identify more than 1000 clone-related problems. To make that easier for Solidity developers to use their solution, They have incorporated it as a web-based application called SmartEmbed in response to developers' comments. Their tool may allow Solidity developers quickly find recurring smart contracts on the live Ethereum blockchain and check their contract against a known set of defects, which can increase users' trust in the contract's dependability. They improve SmartEmbed implementations so they can help developers in real-time for useful applications. Their study has implications for the Ethereum ecosystem and the individual Solidity developer.

Moreover, in this \cite{r3} paper, blockchain security and privacy are described in great depth. They initially describe the concept of blockchains and its utility in the context of online transactions akin to Bitcoin in order to facilitate the conversation. For outlining the core security attributes that are supported as the essential requirements and building blocks for cryptocurrency systems like Bitcoin, they then explore the additional security and privacy qualities that are sought after in many blockchain applications \cite{a2,a3}. The techniques employed in blockchain-based systems to achieve these security attributes are covered at the conclusion, including representative consensus algorithms, hash-chained storage, mixing protocols, anonymous signatures, non-interactive zero-knowledge proof, and others. They contend that this survey will provide readers with a comprehensive understanding of privacy and blockchain security in terms of ideas, attributes, approaches, and systems \cite{a4,a5}.

In order to address more research possibilities, a paper \cite{r4} was published that examined the trend of studies conducted to date and discussed blockchain technology and associated fundamental technologies. Before using blockchain in the cloud computing environment, there are several existing concerns that must be addressed. Even today, blockchain has numerous challenges, including the security of transactions, wallets, and software. Various research have been done to address these problems. User data must be kept confidential and completely deleted when the operation is ended while using blockchain in a cloud-based computing environment. It may be inferred from the data that is still accessible if the individual information is kept and not deleted.
\subsection{Research Goals}
Analysis of previous research and its conclusions, as well as a summary of research efforts into blockchain applications for cyber security \cite{a6,a7}, are the goals of this study. An overview of the questions pursued with a little discussion can be seen in \autoref{researchQuestion}.

\begin{table}[htbp]
\caption{Research Questions}
\label{researchQuestion}
\renewcommand{\arraystretch}{2} 
\begin{tabular}{ p{14em}  p{3.3cm}}
\hline
\textbf{Research Question (RQ)} & \textbf{Discussion}\\ 
\hline
RQ1: What are the most recent studies on platforms and consent protocols for blockchain-enabled smart contracts? & A bunch of studies will be evaluate to figure out what are the major protocols are used in smart contracts and how to smart contract helps to build blockchain. \\

RQ2: What Are the Main Use Cases for Smart Contracts and What Are the Conditions for Using Them? & Different use-cases is evaluate to measure and evaluate the extend of block-chain enabled smart contracts\\

RQ3: What Factors Aid Organizations in Selecting a Blockchain Platform? & Scalability, Ledger Type, Consensus mechanism, programming language and smart contract are evaluate for different blockchain based smart contract to make a selection.\\
\hline
\end{tabular}
\end{table}

\subsection{Contribution and layout}
The contributions provided by this systematic literature review are a combination of past research along with come ongoing tasks are as follow:
\begin{itemize}
  \item IEEE was the top publisher among the top 10 publishers, per an examination of 475 recently released publications.
  \item We looked at 743 publications between 2014 and 2022. Through citation networks created utilising the data gathered from WoS, we have determined the acceptance and authenticity of these research papers.
  \item In order to represent the research, concepts, and considerations in the disciplines of blockchain and smart contracts, we undertake an extensive evaluation of the information available in the group of 21 papers and offer the data.
\end{itemize}

The format of this review paper is as follows: The techniques used to choose the primary studies for analysis in a methodical manner are described in Section 2. The results of all the primary research chosen are presented in Section 3. The findings in relation to the earlier-presented study questions are discussed in Section 4. The research is concluded in Section 5, which also makes some recommendations for additional study.

\section{Research methodology}
We performed the SLR in accordance with the instructions outlined by Kitchenham and Charters [27] in order to accomplish the goal of responding to the research questionnaire. To enable a comprehensive assessment of the SLR, we attempted to progress through study's preparation, executing, and publishing steps in cycles.

\subsection{Selection of primary studies}
By supplying keywords to a particular publication's or search engine's search function, primary research were highlighted. The keywords were chosen to encourage the appearance of study findings that would help answer the research questions. The query terms were: \textit{("smart contracts" OR "smart-contracts" OR "blockchain" OR "block-chain") AND  "security"}

We searched on platforms such as:
\begin{enumerate}
    \item Google Scholar
    \item ACM Digital Library
    \item ScienceDirect
    \item IEEE Xplore Digital Library
\end{enumerate}

Depending on the search platforms, the title, keywords, or abstract were used in the searches. On Nov 7, 2012, we conducted the searches and processed all papers that had been issued up to that point. The inclusion/exclusion criteria, which will be provided in Section 2.2, were used to filter the results from these searches. The criterion enabled us to generate a collection of findings that could subsequently be subjected to Wohlin's \cite{r9} snowballing procedure. Snowballing iterations were performed both forward and backward until no further publications that met the inclusion criteria could be found.

\subsection{Inclusion and exclusion criteria}
With the help of a broad definition of smart contracts and security, we were able to incorporate articles on blockchain technology, Ethereum, cyber security, cryptocurrencies, crypto-transactions, systematic literature reviews, distributed ledgers, Internet of Things, etc. Article titles, keywords, and abstracts were examined to decide if they should be included. The articles' major texts were also carefully examined as needed. More attention was paid to articles that outlined specific parts of the smart contracts that underpin blockchain activities or technology along with its security application. 

Papers providing true facts about implementation of the discussed technology, peer-reviewd articles, and published in a journal or conference proceeding are accepted. Whereas papers relying on financial, commercial or any other out-of-the-topic matters are dismissed. Also, the papers included are only in English language \autoref{inclusionandexclusion} . summarizes the mentioned criterias.

\begin{table}[htbp]
\caption{Inclusion and exclusion criteria for the primary studies.}
\label{inclusionandexclusion}
\renewcommand{\arraystretch}{2.5} 
\begin{tabular}{ p{14em}  p{3.3cm} }
\hline
\textbf{Criteria for inclusion} & \textbf{Criteria for exclusion}\\ 
\hline
The paper must provide actual facts about the execution and application of smart contracts security. & Papers that concentrate on the financial, commercial, or legal implications of blockchain applications \\
The paper must include data about blockchain or comparable distributed ledger systems. & Websites and government papers are examples of irrelevant papers. \\
The article must be a peer-reviewed article that has been published in a conference proceeding or journal. & Papers that are in other language \\
\hline
\end{tabular}
\end{table}

\subsection{Selection results}
\begin{figure}[h]
    \centering
    \captionsetup{justification=centering}
    \includegraphics[width=0.5\textwidth]{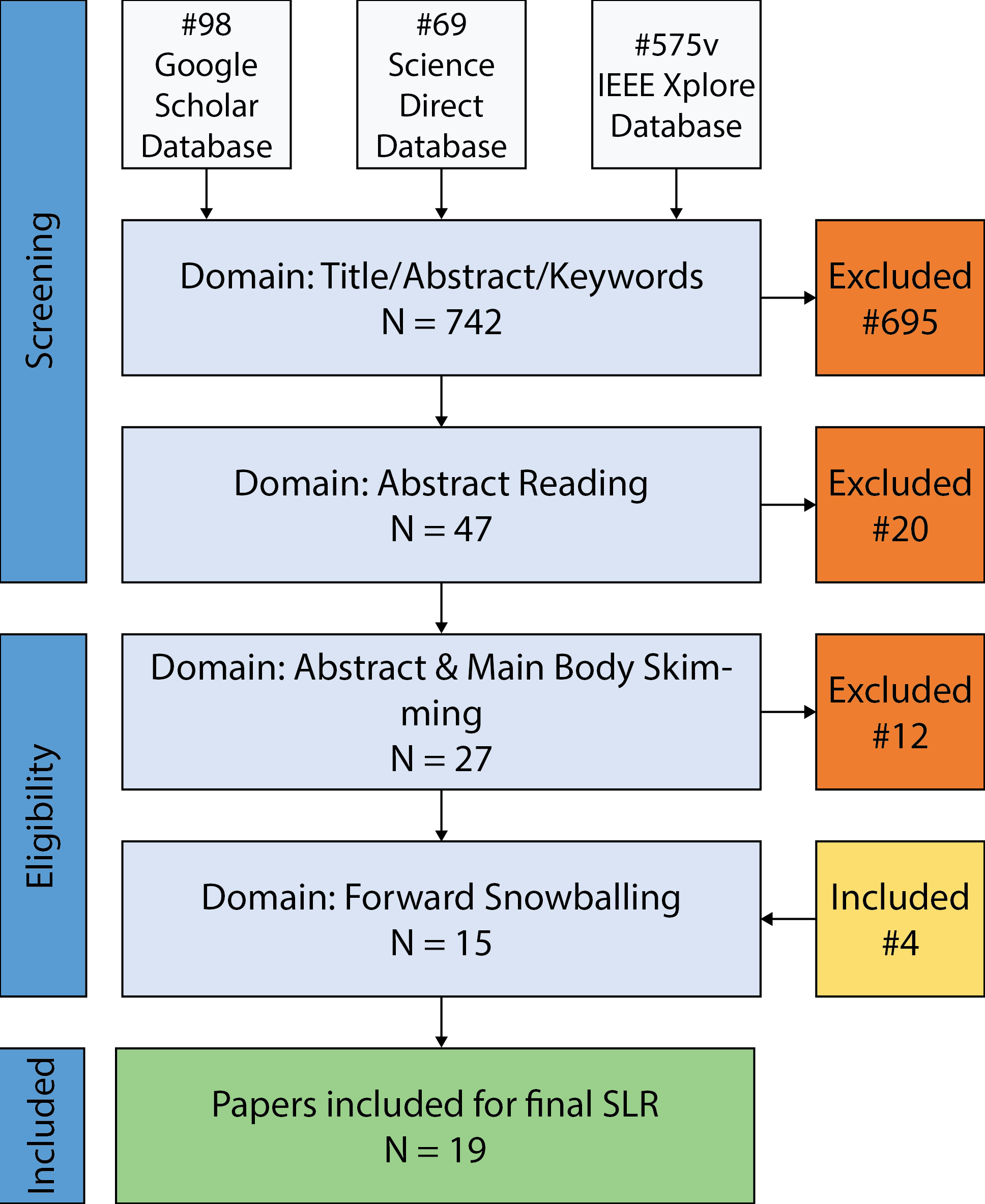}
    \caption{Selection Process}
    \label{fig:selectionProcess}
\end{figure}

\autoref{fig:selectionProcess} displayed the general screening procedures and the order of picking pertinent material. A total of 742 records were discovered in the initial phase (98 from Google Scholar using the sophisticated search approach, 69 from Science Direct, and 575 from IEEE Xplore). The number of literary works was reduced to 47 articles preserved for further title reading after the removal of works of literature like grey literature, extended abstracts, presentations, keynotes, book chapters, non-English language papers, and inaccessible publications. Following that, only 27 articles met the requirements for additional abstract reading. Only 15 articles were left after reading the article abstracts to be read in full. After doing snowballing, 19 of them evaluated smart contracts, and those articles were downloaded for additional screening procedures.

\subsection{Quality assessment}
In accordance with the recommendations provided by Kitchenham and Charters, an evaluation of the main studies' quality was conducted \cite{r5}. This made it possible to evaluate the articles' importance of the research issues while taking any evidence of selection bias and the reliability of observed measurements into account. The evaluation procedure was modelled after the one employed by Hosseini et al. To evaluate their efficacy, four articles were chosen at random and put through following design assessments:
\begin{itemize}
\item
\begin{itemize}
    \item[Stage 1:] \textbf{Smart Contracts.} The article should be based on the implementation of smart contracts or its well-commented deployment to a particular issue.
    \item[Stage 2:] \textbf{Background.} The aims and results of the study must be adequately contextualised. This will make it possible to evaluate the research correctly.
    \item[Stage 3:] \textbf{Application of Smart Contract.} The report must have sufficient information to accurately depict how the solution has been implemented to a particular issue, which will help to address research questions.
    \item[Stage 4:] \textbf{Security and Privacy context.} In order to help in responding to RQ2, the document must explain the security issue.
    \item[Stage 5:] \textbf{Data acquisition.} To assess accuracy, specifics on the data's collection, measurement, and reporting must be provided.
\end{itemize}
\end{itemize}
Excluded papers based on this checklist can be found in \autoref{ExludedStudies}

\begin{table}[htbp]
\caption{Excluded Studies}
\label{ExludedStudies}
\renewcommand{\arraystretch}{2} 
\begin{tabular}{ p{14em}  p{3.3cm}}
\hline
\textbf{Stages of the Criteria Checklist} & \textbf{Excluded Studies}\\ 
\hline
Stage 1: Smart Contracts & \cite{ex6}\cite{ex9}\\
Stage 2: Background & \cite{ex3}\cite{ex7}\cite{ex10}\\
Stage 3: Application of smart contract & \cite{ex1}\cite{ex8}\\
Stage 4: Security and Privacy context & \cite{ex2}\\
Stage 5: Data acquisition & \cite{ex4}\cite{ex5}\cite{ex11}\cite{ex12}\\
\hline
\end{tabular}
\end{table}

\subsection{Data extraction}

Data was then taken from all papers that had passed the quality evaluation in order to evaluate the completeness of the data and verify the accuracy of the information included within the articles. Before being applied to the entire set of studies that have successfully completed the quality evaluation phase, the data extraction technique was first tested on a sample of five studies. Each study's data was taken out, put into categories, and then entered into an excel sheet. The following groups were applied to the information:

\begin{itemize}
  \item Context Data: Data regarding the study's objectives serves as context data.
  \item Qualitative data: The writers' findings and judgments.
  \item Quantitative data: Information gathered through trial and research and used in the study.
\end{itemize}

\subsection{Data analysis}
We gathered the information contained in the qualitative and quantitative data categories in order to achieve the goal of responding to the study questions. We also performed a meta-analysis on the studies that were exposed to the last step of data extraction.\\

\subsubsection{Publication over time}
The term Smart contract was coined by Nick Szabo in 1994. And then an exponential increase can be seen from 2015 year till 2022. The highest trend of publications can be seen in 2017 and 2018 where bitcoin took a hit in the crypto-currency market.\\

\subsubsection{Significant keywords counts}
The most significant keywords used to search and implement the literature review are "smart contract, blockchain, network, transaction, attacks". Other related word queries are distributed kedgers and Internet of Things.

Additionally, publications that addressed the uses of smart contract technology explicitly were chosen for the identification process. Articles that did not include smart contract technology as their main subject were not included, such as those that used the blockchain to explain Bitcoins without mentioning smart contracts. Our collection of references includes papers from year 2006 to 2021.

\section{Finding}
\begin{figure}[h]
    \centering
    \captionsetup{justification=centering}
    \includegraphics[width=0.5\textwidth]{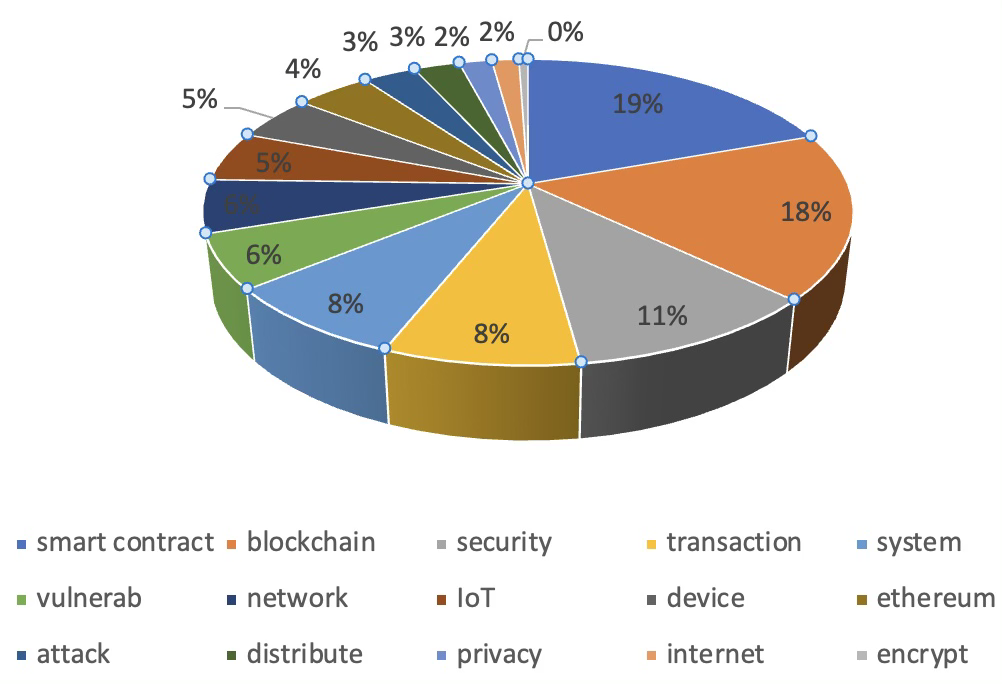}
    \caption{Theme of primary studies}
    \label{fig:keywordweight}
\end{figure}

Each primary research paper was read in full and relevant qualitative
and quantitative data was extracted and summarized in Table 5. All the
primary studies had a focus or theme in relation to how blockchain was
dealing with a particular problem. The focus of each paper is also
recorded below in \autoref{themeofstudy}
The categories found in the main research show that nearly half (47\%) of the papers on blockchain and smart contracts have an interest in IoT device security. With an 18\% rate, transportation and system is the second most popular subject. And the remaining keywords contribute a bit to the original study.This information can be viewed in \autoref{fig:keywordweight}
\begin{table}[htbp]
\caption{Keyword counts in the primary studies}
\label{keywordinstudy}
\renewcommand{\arraystretch}{1.5} 
\begin{tabular}{ p{14em}  p{3.3cm}}
\hline
\textbf{Keywords} & \textbf{Counts}\\ 
\hline
smart contracts & 1283 \\
blockchain & 978 \\
security & 623 \\
transaction & 455 \\
system & 447 \\
vulnerable & 318 \\
network & 311 \\
IoT & 294 \\
device & 266 \\
ethereum & 248 \\
attack & 175 \\
distribute & 151 \\
privacy & 108 \\
internet & 89 \\
encrypt & 30\\
\hline
\end{tabular}
\end{table}

\begin{table}[htbp]
\caption{The key research' main results and topics}
\label{themeofstudy}
\setlength{\tabcolsep}{5pt} 
\renewcommand{\arraystretch}{2.5} 
\begin{tabular}{|p{1cm} p{5cm} p{1.5cm}|}
\hline
\textbf{Primary Study} & \textbf{Key Qualitative and Quantitative Data Reported}& \textbf{Types of Security Applications}\\ 
\hline
\cite{ps1} & There are many security vulnerabilities and bugs in the smart contract due to inconsistencies in the design of smart contracts. They proposed a self-adaptive security modeling approach for modeling and analyzing Service Level Agreement (SLA) based smart contracts using attack scenarios and goal models. & Blockchain-Based Cloud (BBC) \\
\cite{ps2} & They introduced a protocol that requires secure device deployment, communication, monitoring, and upkeep. They also suggested using smart contracts to create and manage general IoT activities. By understanding tasks that are sent to the devices through smart contract transactions, this architecture enables the gadgets to operate independently. & IoT Security \\
\cite{ps3} & Smart contracts are distributed database ledger-based, event-driven, full-state systems. This study examines the state of research on smart contracts and its difficulties in terms of security, scalability, and maintainability. To fulfil the terms of the contract, smart contracts can be automatically applied to computer programmes operating on the blockchain. & Blockchain \\
\cite{ps4} & They analyse the provided use cases to determine the possible features for expanding the coverage of smart contracts in security token offerings (STO) before designing the operational model of security tokens. They also suggest a smart contract-based security token management system that integrates with several blockchain ledgers and is built on an expanded smart contract framework. & Security Token Offerings (STO) \\
\cite{ps5} & Numerous flaws, including re-entrancy, transaction origin, call stack depth exception, timestamp reliance, and transaction-ordering dependence, were found. In addition, they suggested a few methods to deal with the issues, like ZeppelinOS, SolCover, HackThisContract, Security audits, etc. & Blockchain \\
\cite{ps6} & Hyperledger Fabric's smart contracts run in a docker virtual machine, and do not need to consume Gas. The solution they propose can help users apply for social insurance, welfare and minimum living guarantee at home during the epidemic. Smart contracts and their algorithms are introduced and studied in this article. & Peer-to-peer computing \\

\hline
\end{tabular}
\end{table}

\begin{table}[htbp]
\caption{\autoref{themeofstudy} (continued)}
\setlength{\tabcolsep}{5pt} 
\renewcommand{\arraystretch}{2.5} 
\begin{tabular}{|p{1cm} p{5cm}  p{1.5cm}|}
\hline
\textbf{Primary Study} & \textbf{Key Qualitative and Quantitative Data Reported}& \textbf{Types of Security Applications}\\ 
\hline
\cite{ps7} & To provide a secure system for IoT devices in Home automation systems, authentication strategy that integrates attribute-based access control utilising smart contracts with ERC-20 Token (Ethereum Request For Comments) and edge computing. Through the offloading of more demanding compute jobs, edge servers help the system scale. & IOT(Spcifically for Smart Home) \\
\cite{ps8} & It provided a brand-new, two-phase structure built on trustworthy Intel SGX technology. They create a pre-execution system for smart contracts stored in unreliable memory. They produce a small read-write set and a Merkle Forest data structure. At end, it incorporate every technique suggested in into the open-source BFT-SMaRt system. & Blockchain Using SGX  \\
\cite{ps9} & Use solidity, which has shown to be more secure than md5 and SHA, to verify each & Peer-to-peer connection(for transaction) \\
\cite{ps10} & It ensures the immutability and transparency of energy transactions using the block chain, creates ERC20 tokens based on smart contracts, and executes transactions automatically without such involvement of a third party. Transactions may also be expanded to multiple transaction circumstances. The Energy Storage System (ESS), to which both the seller and the buyer belong, is used to transmit the energy during the transaction. & Peer-to-peer connection \\
\cite{ps11} & To develop the smart contract, use the Metamask wallet and the Solidity programming language. The outcomes of the Ropsten blockchain network's smart contract implementation are then evaluated and contrasted with comparable efforts. The research shows that the suggested framework has improved security and privacy. & Blockchain (for every trading) \\
\cite{ps12} & IoT devices have a direct connection to the blockchain, and a smart contract that controls them manages changes to their ownership, pin, and other data. which prevents the outsider from interfering They also offer guidelines for these IoT gadgets and the smart contracts that manage them. & IoT network \\
\cite{ps13} & To prevent Man-In-The-Middle (MITM) attempts on multifactor authentication, a blockchain-based two-factor authentication technique for web-based access to sensor data can be employed. The suggested approach uses the Ethereum blockchain and smart contracts technology to deliver a quick and user-focused authentication. & Two factor authentication (using Blockchain) \\
\hline
\end{tabular}
\end{table}

\begin{table}[htbp]
\caption{\autoref{themeofstudy} (continued)}
\setlength{\tabcolsep}{10pt} 
\renewcommand{\arraystretch}{2.5} 
\begin{tabular}{|p{1cm} p{4cm} p{1.5cm}|}
\hline
\textbf{Primary Study} & \textbf{Key Qualitative and Quantitative Data Reported}& \textbf{Types of Security Applications}\\ 
\hline
\cite{ps14} & The existing functional requirements and complexity of the programme code for the smart contract are taken into account when recommending a five-step smart contract audit plan. This plan consist of Agreement on specification, Manual code review, Testing, Automated code analysis and  audit report. & Blockchain \\
\cite{ps15} & As stated in the article, Software-Defined Networking (SDN) may be utilised to prevent unauthorised access and DoS attempts. Based on the use cases in several sectors, Hyperledger Fabric is the foundation of this. & Blockchain \\
\cite{ps16} & Smart contract flaws may be found with SoliAudit. For fuzz testing with anomalous analysis, it automatically builds a fuzzer contract that successfully analyses real-world and CTF contract instances. By the experiment, the dynamic fuzzer has the ability to find undiscovered vulnerabilities with accuracy unto 90\% because it focuses on reentrancy and arithmetic flaws. & Security (smart contract) \\
\cite{ps17} & SASC is a static analysis tool that can create an invocation connection topology diagram and identify potential logic hazards. They evaluated 2,952 decentralized applications, and the findings of the trial demonstrated how simple and efficient this technology is. Despite the fact that they are able to identify possible hazards, risks are not actually errors. & Cryptocurrency \\
\cite{ps18} & By upload the encrypted information related to the movie to the Blockchain network; on the normal server side, employ smart contracts to control the processes of embedding and extracting the encrypted information that has been included in the digital video. Then, confirm, track, and preserve the private data that is saved in the Blockchain network and the actual digital video files, respectively. It has been demonstrated that the suggested architecture might provide accurate and highly effective digital video security protection. & Data Hiding \\
\cite{ps19} & Utilizing the inherent security mechanisms of the blockchain, blockchain-based implementation processing system suggests using smart contracts to automate the many procedures needed in the validation and verification of applications. & Blockchain(for education) \\
\hline
\end{tabular}
\end{table}

\section{Discussion}
Smart contract usability is impacted by a number of variables, including data transmission rate, information update rate, and domain-specific needs. Clarifying the application environment for smart contracts is crucial for their development and planning. Preliminary keyphrases reveal that there are a large number of studies on smart contracts. Smart contracts and genuinely distributed decentralised systems technologies have been created for only 10 years and are obviously still in their development. A significant number of the major studies chosen are experimental recommendations or notions for solving today's challenges, with little quantitative data and few actual implementations.
Gateway flaws, secret keys security issues, blockchain integration systems, absence of full-scale testing, a lack of rules and regulations, unproven code, and smart contract flaws are among the most prevalent issues .Both illegal miners and consumers can take advantage of certain kinds of vulnerabilities, claim the authors at \cite{r10}. Several researchers have concentrated on studying the most frequent mistakes in smart contracts and attempted to fix them in order to enhance the creation of smart contracts secure \cite{r11}, \cite{r12}. Recent publications \cite{r13} present techniques for static code analysis vulnerability detection. All verified smart contracts are made to adhere to a guidelines by Quantumstamp. The decentralised security mechanism they built enhances the blockchain architecture.
\\

\textbf{RQ1: What are the most recent studies on platforms and consent protocols for blockchain-enabled smart contracts?}\\

\textbf{Current Research on Smart Contracts}
In January 2009, Satoshi Nakamoto created the bitcoin blockchain. Both the actual evidence of smart contracts as well as the decentralized peer-to-peer digital money Bitcoin were presented in his study \cite{r14}. Those two essential ideas provide the basis for the majority of the SLR results that follow and have substantially influenced the development of blockchain technology. Since then, the emphasis has migrated to other fields than economics since it may help firms assure integrity, boost efficiency, and cut down on redundancies \cite{r15}, \cite{r16}. Implementing smart contracts may be highly difficult, particularly for non-experts \cite{ps20}. Therefore, it is essential to comprehend the speed and scalability constraints of smart contract functionalities.

\textbf{Platforms for Smart Contracts}
Various blockchain systems allow for the development and processing of smart contracts, based on a number of factors and traits \cite{ps21}. In this part, we identified several crucial technical characteristics of the five systems that received the most citations throughout the 30 publications we analysed. In light of the kind of enterprise, database, smart contract capability, transaction costs, accessible languages, consensus process, and administration, we emphasized the significant distinctions between these platforms.

\begin{enumerate}
\item Bitcoin
A decentralised digital money network is called Bitcoin. It makes use of a permissionless blockchain network to provide an open and permanent record of all monetary transactions. To create 256-bit long hashes for documents that can be used confirm the validity, Bitcoin utilises the cryptographic hash algorithm SHA256 \cite{r17}. The use of Bitcoin is severely constrained by the proof-of-work consensus process that it depends on. The fresh chain's block is produced by nodes inside a bitcoin network by solving an algorithmic puzzle in parallel.

\item Ethereum
Created in July 2015, Ethereum is a decentralised online system for financial transactions as well as other uses. Ethereum is a programmable platform that allows for the compilation and implementation of payment systems in a variety of languages, unlike numerous other blockchains \cite{ps21}. In fact, Ethereum offers the Ethereum Virtual Computer (EVM), a Turing-complete machinery that allows the execution of numerous programming languages. The most well-known ones are Solidity and Vyper, which are mostly utilised in the creation of complicated smart contracts \cite{ps21}. Ethereum has implemented the Proof-of-Work agreement technique to verify its calculations, following the lead set by Bitcoin.

\item Hyperledger Fabric
The Linux Foundation has created an open-source, decentralized distributed ledger known as Hyperledger Fabric. Extensive customization of the consensus process and programming language makes Fabric one of the most modular and flexible systems \cite{r18}. Hyperledger Fabric is the first blockchain platform to support general-purpose programming languages such as Python, Go, Java, JavaScript, and Node.js, using a plugin consensus framework to customize for specific use cases. Scalability and performance issues are other issues Fabric is known to address.
\end{enumerate}
\textbf{Programming Languages for Smart Contracts}

The development of smart contracts on the blockchain is still in its infancy. As a result, new programming languages are being developed in accordance with the architecture of each platform. In fact, the most popular programming languages for smart contracts are emphasised in this article since it is essential to see which ones are reinforced by whatever blockchain stage before starting any project. Due to the intricacy of their contracts, we decide to concentrate on these four languages. There are four major languages seen as solidity, viper, rholang and kotlin, from which two are explained here:

\begin{enumerate}
\item Vyper
The programming language Vyper was developed to fend off errors and assaults \cite{ps23}. It is closely related to Serpent language and is descended from Python. Due to Python's high-level syntax, Vyper offers more efficiency and trustworthy outcomes compared to Solidity.

\item Rholang
A concurrent programming language with behavioural typing, Rholang is officially patterned after Rho-calculus. Rchain blockchain \cite{ps23} was the first software to use this programming language.
\end{enumerate}

\textbf{Consensus Tools in Smart Contracts Powered by Blockchain}
The consensus protocols increase the proper and effective implementation and execution of a smart contract. In actuality, a network's transactions should all be recorded, and any relevant smart contracts should be carried out. The nodes of the same network perform these two activities in a unified and predictable manner. Nodes should first come to consensus in order to achieve this state.

Recently, several consensus protocols were introduced. However, Proof-of-Work (PoW) and Proof-of-Stake (PoS) are the most popular ones.

\begin{enumerate}
    
\item Proof-of-Work (PoW)
Every block of blockchains includes information that has been firmly recorded. Cryptography is a method for creating trust. Miners must complete a proof-of-work by resolving a mathematical puzzle in order for the network's members to produce and validate a block . \autoref{fig:proofofwork} depicts the PoW protocol's flow.
\begin{figure}[h]
    \centering
    \captionsetup{justification=centering}
    \includegraphics[width=0.5\textwidth]{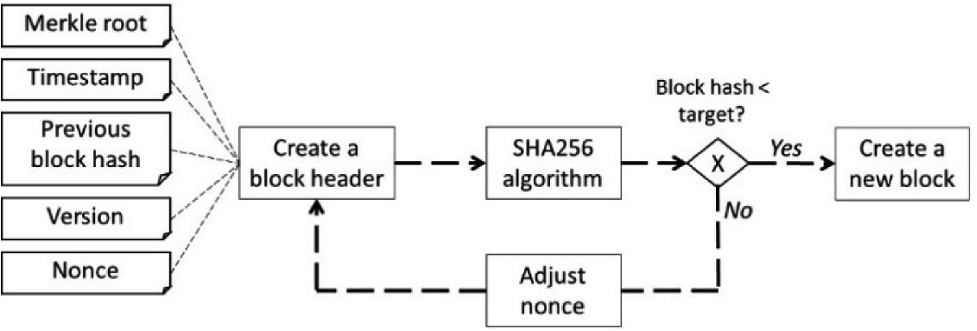}
    \caption{Proof of work \cite{r19}}
    \label{fig:proofofwork}
\end{figure}

\item Proof-of-Stake(PoS)
A network can employ the PoS method to reach distributed consensus without the energy loss of PoW. PoS picks the participants that will build the next block depending on how wealthy they are, in contrast to PoW's rewarding mechanism for coin-miners, which is grounded on completing challenging problems and systems. \autoref{fig:proofofstack} depicts the PoS protocol's flow.
\begin{figure}[h]
    \centering
    \captionsetup{justification=centering}
    \includegraphics[width=0.5\textwidth]{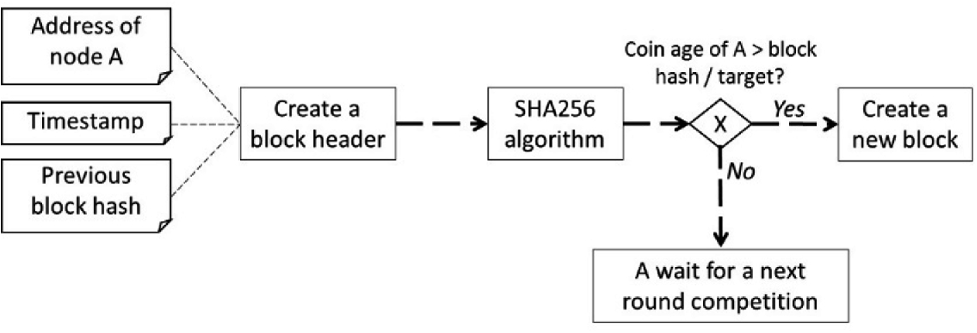}
    \caption{Proof of stake \cite{r19}}
    \label{fig:proofofstack}
\end{figure}

\end{enumerate}

\textbf{RQ2: What Are the Main Use Cases for Smart Contracts and What Are the Conditions for Using Them?}\\
We concentrated our research and analysis based on the use cases and goals of smart contracts in order to quantify and assess the level of business value provided by blockchain-enabled smart contracts.

Use of smart contracts powered by blockchain has spread to a variety of industries. The three key drivers behind this technology adoption are data protection, trust, and accountability \cite{r20,a7,a8}. But it could also be used for other things in some places.

To secure not one data confidentiality and confidence but also transparency and contaminate material, major application domains including healthcare, voting, the pharmaceutics, and the schooling institution have implemented the block chain technology smart contracts. The same goals of smart contracts are shared by IoT and data security \cite{r18}.

The deployment of blockchain-enabled smart agreements in Smart City applications \cite{r20}, management of occupational processes, as well as land registration and land is a consequence of the requirement for trust-based transactions. Data relevance is another desired feature that the market forecast \cite{r20} discovered in blockchain-enabled payment systems.

Other application domains need efficiency, security, and efficiency. Relevant examples of these sectors include industrial output, energy supplies \cite{r18}, management of supply chains, and financial \cite{r18} \cite{r20,a10}.

We provide some pertinent instances of platforms for each application area based on the findings in Table 2. After comparing the key features of the public blockchain with the needs for the domain, the platform was selected. Although Ethereum continues to be the most popular platform owing to its high information immutableness , it still suffers from major performance and scalability issues, making it increasingly probable that alternative platforms will take its place.

NXT \cite{ps24}, for instance, intends to integrate security and provide efficiency in order to prevent end-to-end deferrals in the financial sphere. WAVES  can be used by application sectors that want to achieve great result in terms of cost and time savings since it also increases scalability and speed.

Cross-industry platforms like R3 Corda and EOS \cite{ps24} promote confidence and transparency among the network's many participants. They are suitable platforms for the supply chain application domain because of their characteristics.

The secrecy and security of records are a key emphasis of the Quorum and Hyperledger Fabric platforms. They work with apps that demand swift private transactions, which are crucial for patients and other users of the healthcare system.

\textbf {RQ3: What Factors Aid Organizations in Selecting a Blockchain Platform?}\\
Practically speaking, we provide a grid of criteria that businesses may use to select the best public blockchain for their operations. We defined five key technological characteristics and requirements for the platform based on the research that the organisation should support in order to meet its needs.
\begin{enumerate}

\item Scalability
Application of smart contracts has major challenges in terms of scalability \cite{r16}. In fact, due to their transaction-intensive nature, several application areas, like IoT, demand high resilience and scalability \cite{ps24}. Data storage on the blockchain might lead to serious scalability problems [sl06]. An organisation must select a public blockchain that can expand to accommodate expansion for this reason.

\item Ledger Type
Blockchain, a young technology, offers three types of ledgers: consortia, private, and public. The network scope determines which ledger category to use. For instance, anybody can be a lump on public networks. In grouping networks, nodes are assigned and authorizations are regulated. Permissions are more tightly controlled in private networks, which leads to very little decentralisation. Since there are several variations of blockchain, not all of its systems offer completely open networks or less decentralised ledgers, like R3 Corda, which would be exclusively permissioned.

\item The consensus process
Some platforms' usability is constrained by non-adaptable consensus protocol\cite{r16}te, and an appropriate consensus procedure must provide security and offer accountability tolerance . It is well recognised that PoW uses a lot of energy and has a very low throughput of only 3–7 transactions per second. There are various protocols and approaches that may be used to reduce the restriction of this method, such as Merkle tree \cite{r21}, to address the scalability problem for systems that only allow PoW, such as Ethereum. Platforms based on PoS and DPoS may also be an useful substitute.

\item Programming dialect
The advent of the blockchain has led to the introduction of several programming languages \cite{a9}. The most well-known of them is Solidity, a language created expressly for blockchain that was strongly impacted on JavaScript. As a result, a company needs to find out which programming languages a blockchain platform supports. In addition to the four language categories mentioned above, functional, procedural, declarative, and object-oriented languages (such as C++ and Python) were also discovered.

\item Smart contracts support
Some blockchain platforms might not support smart contracts, which are in charge of carrying out actions carried out by ordinary programming languages in a blockchain network. Quorum \cite{r15} is an illustration of this type of platform.
\end{enumerate}
These requirements are not all included in this list. The simplicity of use, toolchain maturity, and people resources and capabilities are just a few other variables that might influence the selection of the best distributed platform. The five previously stated criteria, however, are the only ones that this study article focuses on.

\section{Future research directions of smart contracta security}
In considering all the options, it is important to keep in mind that blockchain technology remains in its infancy so it will take a while and development before it enters the public. Given that a smart contract is really a "contract" that is subject to strict restrictions, the regulatory components of the contract need also be taken good care of.

Some nations still have legal frameworks from the Eighteenth Century that are over about 140 years old. Since there is no one body gathering information from the blockchain, it is possible that data security regulations and the associated consequences for not complying with them may not be effectively followed.

Decentralization may be wonderful, but some purists may ignore the worry of having a centralised authority to hold responsible. There is a potential that someone with superior technological understanding might make flaws in the shared ledger directly, which could lead to the loss of information, revenue, confidence, and ethics.

Nevertheless, people are becoming more knowledgeable about blockchain as well as its potential. Smart contracts are evolving to reach a delicate balance between conventional ideals and contemporary technologies. We may anticipate smart contracts influencing, if not controlling, each aspect of our life that is tied to the word "contract," once both of them are in existence and yet when they eventually merge.

\section{Conclusion}
This research undertook a methodical content analysis that outlined the key characteristics of block chain technology consensus mechanism and the current state of the art in its many uses. We analysed a wide range of scientific details and standards, including the accepted computer languages and agreement processes, to showcase a large number of network platforms.

We have a tendency to think that such a research will assist corporations in comprehending their demands and specifications for the creation of their smart contracts apps. Indeed, not all blockchain platforms are appropriate for all networks. We came to the conclusion that one of the most important things an organisation should understand about its execution environment are: (i) check to see if the system deals with blockchain networks; (ii) verify the consensus protocols aided by this system; (iii) know what computer scripts, the Software Development Kits (SDKs) of the runtime environments; and (iv) exactly what sort of scalability would the solution require. The firm will be able to select the best blockchain - based platform thanks to this early diagnosis, which will also assist to prevent the serious technical problems that can arise in terms of speed and scale when an agreement is implemented. The discussion section's little used practical example brought attention to the value of our study on smart contracts systems. In order to enhance the system grid and stay consistent with the applications areas' changing needs, future studies can broaden the original study focus and investigate additional determining criteria.

\textbf{Declarations of interest}
\\ 
There is no conflict of interest.
\\
\textbf{Acknowledgement} 
\\
None.

\textbf{Primary Studies} 

\vspace{12pt}
\color{red}


\begin{thebibliography}{00}
\bibitem{r1} J. M. Montes, C. E. Ramirez, M. C. Gutierrez, and V. M. Larios, ‘Smart Contracts for supply chain applicable to Smart Cities daily operations’, in 2019 IEEE International Smart Cities Conference (ISC2), 2019, pp. 565–570.
\bibitem{a1} Yazdinejad, Abbas, et al. "An energy-efficient SDN controller architecture for IoT networks with blockchain-based security." IEEE Transactions on Services Computing 13.4 (2020): 625-638.
\bibitem{r2} Z. Gao, ‘When Deep Learning Meets Smart Contracts’, in Proceedings of the 35th IEEE/ACM International Conference on Automated Software Engineering, Virtual Event, Australia, 2020, pp. 1400–1402.
\bibitem{a2} Yazdinejad, Abbas, et al. "Decentralized authentication of distributed patients in hospital networks using blockchain." IEEE journal of biomedical and health informatics 24.8 (2020): 2146-2156.
\bibitem{r3} R. Zhang, R. Xue, and L. Liu, ‘Security and Privacy on Blockchain’, ACM Comput. Surv., vol. 52, no. 3, Jul. 2019.
\bibitem{r4} J. H. Park and J. H. Park, ‘Blockchain Security in Cloud Computing: Use Cases, Challenges, and Solutions’, Symmetry, vol. 9, no. 8, 2017.
\bibitem{r5}
B. Kitchenham and S. Charters, ‘Guidelines for performing systematic literature reviews in software engineering’, 2007.
\bibitem{a3}Yazdinejad, Abbas, et al. "Blockchain-enabled authentication handover with efficient privacy protection in SDN-based 5G networks." IEEE Transactions on Network Science and Engineering 8.2 (2019): 1120-1132.
\bibitem{r6} E. Insurance and O. P. Authority, ‘Discussion Paper on blockchain and smart contracts in insurance’. 2021.
\bibitem{r9} C. Wohlin, ‘Guidelines for snowballing in systematic literature studies and a replication in software engineering’, in Proceedings of the 18th international conference on evaluation and assessment in software engineering, 2014, pp. 1–10.
\bibitem{r10} L. Loi, C. Duc-Hiep, O. Hrishi, P., S. and A. Hobor, "Making Smart Contracts Smarter", ACM SIGSAC Conferece on Computer and Communications Security (CCS '16), pp. 254-269, 2016.
\bibitem{a4} Yazdinejad, Abbas, et al. "Enabling drones in the internet of things with decentralized blockchain-based security." IEEE Internet of Things Journal 8.8 (2020): 6406-6415.
\bibitem{r11}K. Bhargavan, A. Delignat-Lavdaoud, C. Fournet, A. Gollamudi, G. Gonthier, N. Kobeissi, et al., "Formal verification of smart contracts: Short Paper", ACM Workshop on Programming Languages and Analysis for Security (PLAS 16), 2016.
\bibitem{r12} J. Pettersson and R. Edstorm, Safer smart contracts through type-driven development: using dependent and polymorphic types for safer development of smart contracts, 2016.
\bibitem{a5} Yazdinejad, Abbas, et al. "Block Hunter: Federated Learning for Cyber Threat Hunting in Blockchain-based IIoT Networks." IEEE Transactions on Industrial Informatics (2022).
\bibitem{r13} N. Atzei, M. Bartoletti and T. Cimoli, "A survey of attacks on Ethereum smart contracts", IACR Cryptology ePrint Archive, pp. 99-110, 2016.
\bibitem{r14} Buterin, V.:  A next-generation  smart contract  and decentralized application platform. Ethereum White Paper, 2014.
\bibitem{a6} Yazdinejad, A., Dehghantanha, A., Parizi, R. M., Srivastava, G., \& Karimipour, H. (2023). Secure Intelligent Fuzzy Blockchain Framework: Effective Threat Detection in IoT Networks. Computers in Industry, 144, 103801.
\bibitem{r15} Rouhani, S., Deters, R.: Security, Performance, and Applications of Smart Contracts: A Systematic Survey. IEEE Access. 7, 50759–50779 (2019).
\bibitem{r16} Udokwu, C.,  Kormiltsyn, A.,  Thangalimodzi, K., Norta,  A.: The  State of  the Art for Blockchain-Enabled Smart-Contract Applications in the Organization. In: 2018 Ivanni-kov Ispras Open Conference (ISPRAS). 137–144 (2018).
\bibitem{a7}Rabieinejad, Elnaz, Abbas Yazdinejad, and Reza M. Parizi. "A deep learning model for threat hunting in ethereum blockchain." 2021 IEEE 20th International Conference on Trust, Security and Privacy in Computing and Communications (TrustCom). IEEE, 2021.
\bibitem{r17} Stefanović, M.,  Ristić, S., Stefanović, D., Bojkić, M., Pržulj,  D.: Possible Applications of Smart Contracts in Land Administration. In: 2018 26th Telecommunications Forum. 420–425 (2018).
\bibitem{a8}Yazdinejad, Abbas, et al. "Energy efficient decentralized authentication in internet of underwater things using blockchain." 2019 IEEE Globecom Workshops (GC Wkshps). IEEE, 2019.
\bibitem{r18} Wang, S.,  Ouyang, L., Yuan,  Y., Ni,  X., Han,  X., Wang, F.-Y.: Blockchain-Enabled Smart Contracts: Architecture, Applications, and Future Trends. IEEE Transactions on Systems, Man, and Cybernetics: Systems. 49, 2266–2277 (2019).
\bibitem{r19}Zhang and Lee, 2019 (Zhang, S., Lee, J.-H.: Analysis of the main consensus protocols of blockchain. ICT Express. Online first (2019). 
https://doi.org/10.1016/j.icte.2019.08.001)
\bibitem{a9}Yazdinejad, Abbas, et al. "P4-to-blockchain: A secure blockchain-enabled packet parser for software defined networking." Computers \& Security 88 (2020): 101629.
\bibitem{r20} Tinu, N. S.: Ethereum: A Blockchain Platform with smart contract support for Distrib-uted Application  Development, INTERNATIONAL  JOURNAL OF  INFORMATION AND COMPUTING SCIENCE, 6(7), 142-145, available at http://ijics.com/ (2019)
\bibitem{r21}Kim, S., Kwon, Y., Cho, S.: A survey of scalability solutions on blockchain. In: 2018 International Conference on Information and Communication Technology Convergence (ICTC), pp. 1204–1207. IEEE (2018). https://doi.org/10.1109/ICTC.2018.8539529
\bibitem{a10}Kazemi, Mostafa, and Abbas Yazdinejad. "Towards automated benchmark support for multi-blockchain interoperability-facilitating platforms." arXiv preprint arXiv:2103.03866 (2021).
\\
\end{thebibliography}

\begin{thebibliography}{00}
\bibitem{ps1} I. Singh and S.-W. Lee, ‘Self-Adaptive Security for SLA Based Smart Contract’, in 2021 IEEE 29th International Requirements Engineering Conference Workshops (REW), 2021, pp. 388–393.
\bibitem{ps2}J. Wickström, M. Westerlund, and G. Pulkkis, ‘Smart Contract based Distributed IoT Security: A Protocol for Autonomous Device Management’, in 2021 IEEE/ACM 21st International Symposium on Cluster, Cloud and Internet Computing (CCGrid), 2021, pp. 776–781.
\bibitem{ps3}J. Dongfang and L. Wang, ‘Research on smart contract technology based on block chain’, in 2022 International Conference on Artificial Intelligence in Everything (AIE), 2022, pp. 664–668.
\bibitem{ps4}S. Fujimoto and K. Omote, ‘Proposal of a smart contract-based security token management system’, in 2022 IEEE International Conference on Blockchain (Blockchain), 2022, pp. 419–426.
\bibitem{ps5}A. Dika and M. Nowostawski, ‘Security Vulnerabilities in Ethereum Smart Contracts’, in 2018 IEEE International Conference on Internet of Things (iThings) and IEEE Green Computing and Communications (GreenCom) and IEEE Cyber, Physical and Social Computing (CPSCom) and IEEE Smart Data (SmartData), 2018, pp. 955–962.
\bibitem{ps6}S. Tang, Z. Wang, J. Dong, and Y. Ma, ‘Blockchain-Enabled Social Security Services Using Smart Contracts’, IEEE Access, vol. 10, pp. 73857–73870, 2022.
\bibitem{ps7}A. Qashlan, P. Nanda, and X. He, ‘Security and Privacy Implementation in Smart Home: Attributes Based Access Control and Smart Contracts’, in 2020 IEEE 19th International Conference on Trust, Security and Privacy in Computing and Communications (TrustCom), 2020, pp. 951–958.
\bibitem{ps8}M. Fang, Z. Zhang, C. Jin, and A. Zhou, ‘High-Performance Smart Contracts Concurrent Execution for Permissioned Blockchain Using SGX’, in 2021 IEEE 37th International Conference on Data Engineering (ICDE), 2021, pp. 1907–1912.
\bibitem{ps9}P. Khandelwal, R. Johari, V. Gaur, and D. Vashisth, ‘BlockChain Technology based Smart Contract Agreement on REMIX IDE’, in 2021 8th International Conference on Signal Processing and Integrated Networks (SPIN), 2021, pp. 938–942.
\bibitem{ps10}S. J. Pee, E. S. Kang, J. G. Song, and J. W. Jang, ‘Blockchain based smart energy trading platform using smart contract’, in 2019 International Conference on Artificial Intelligence in Information and Communication (ICAIIC), 2019, pp. 322–325.
\bibitem{ps11}M. Nazari, S. Khorsandi, and J. Babaki, ‘Security and Privacy Smart Contract Architecture for Energy Trading based on Blockchains’, in 2021 29th Iranian Conference on Electrical Engineering (ICEE), 2021, pp. 596–600.
\bibitem{ps12}M. Shurman, A. A.-R. Obeidat, and S. A.-D. Al-Shurman, ‘Blockchain and Smart Contract for IoT’, in 2020 11th International Conference on Information and Communication Systems (ICICS), 2020, pp. 361–366.
\bibitem{ps13}M. Abubakar, Z. Jaroucheh, A. Al Dubai, and X. Liu, ‘A Lightweight and User-centric Two-factor Authentication Mechanism for IoT Based on Blockchain and Smart Contract’, in 2022 2nd International Conference of Smart Systems and Emerging Technologies (SMARTTECH), 2022, pp. 91–96.
\bibitem{ps14}I. Popchev, I. Radeva, and V. Velichkova, ‘Auditing blockchain smart contracts’, in 2022 International Conference Automatics and Informatics (ICAI), 2022, pp. 276–281.
\bibitem{ps15}M. Almakhour, A. Wehby, L. Sliman, A. E. Samhat, and A. Mellouk, ‘Smart Contract Based Solution for Secure Distributed SDN’, in 2021 11th IFIP International Conference on New Technologies, Mobility and Security (NTMS), 2021, pp. 1–6.
\bibitem{ps16}J.-W. Liao, T.-T. Tsai, C.-K. He, and C.-W. Tien, ‘SoliAudit: Smart Contract Vulnerability Assessment Based on Machine Learning and Fuzz Testing’, in 2019 Sixth International Conference on Internet of Things: Systems, Management and Security (IOTSMS), 2019, pp. 458–465.
\bibitem{ps17}E. Zhou et al., ‘Security Assurance for Smart Contract’, in 2018 9th IFIP International Conference on New Technologies, Mobility and Security (NTMS), 2018, pp. 1–5.
\bibitem{ps18}H. Zhao, Y. Liu, Y. Wang, and Y. Huang, ‘Hiding Data into Blockchain-based Digital Video for Security Protection’, in 2020 3rd International Conference on Smart BlockChain (SmartBlock), 2020, pp. 23–28.
\bibitem{ps19}R. J. Kutty and N. Javed, ‘Secure Blockchain for Admission Processing in Educational Institutions’, in 2021 International Conference on Computer Communication and Informatics (ICCCI), 2021, pp. 1–4.
\bibitem{ps20} Weingaertner,  T., Rao,  R.,  Ettlin, J.,  Suter,  P., Dublanc,  P.:  Smart Contracts  Using Blockly:  Representing a  Purchase Agreement  Using  a  Graphical Programming  Lan-guage. In: 2018 Crypto Valley Conference on Blockchain Technology (CVCBT). 55–64 (2018).
\bibitem{ps21} Wang, S., Yuan, Y., Wang, X., Li, J., Qin, R., Wang, F.-Y.: An Overview of Smart Con-tract: Architecture, Applications, and Future Trends. In: 2018 IEEE Intelligent Vehicles Symposium (IV). 108–113 (2018).
\bibitem[ps22]Zheng, Z., Xie, S., Dai, H.-N., Chen, W., Chen, X., Weng, J., Imran, M.: An overview on smart contracts: Challenges, advances and platforms. Future Generation  Computer Systems. 105, 475–491 (2020)
\bibitem{ps23}Knecht,  M.: Mandala:  A Smart  Contract Programming  Language. arXiv:1911.11376 [cs]. (2019).
\bibitem{ps24}Rashid, A., Siddique, M.J.: Smart Contracts Integration between Blockchain and Internet of Things: Opportunities and Challenges. In: 2019 2nd International Conference on Ad-vancements in Computational Sciences (ICACS). 1–9 (2019). 


\bibitem{ex1}J. Liu and Z. Liu, ‘A Survey on Security Verification of Blockchain Smart Contracts’, IEEE Access, vol. 7, pp. 77894–77904, 2019.
\bibitem{ex2}Z. Wan, X. Xia, D. Lo, J. Chen, X. Luo, and X. Yang, ‘Smart Contract Security: A Practitioners’ Perspective’, in 2021 IEEE/ACM 43rd International Conference on Software Engineering (ICSE), 2021, pp. 1410–1422.
\bibitem{ex3}M. Demir, M. Alalfi, O. Turetken, and A. Ferworn, ‘Security Smells in Smart Contracts’, in 2019 IEEE 19th International Conference on Software Quality, Reliability and Security Companion (QRS-C), 2019, pp. 442–449.
\bibitem{ex4}R. Pise and S. Patil, ‘A Deep Dive into Blockchain-based Smart Contract-specific Security Vulnerabilities’, in 2022 IEEE International Conference on Blockchain and Distributed Systems Security (ICBDS), 2022, pp. 1–6.
\bibitem{ex5}J. Chen, ‘Finding Ethereum Smart Contracts Security Issues by Comparing History Versions’, in 2020 35th IEEE/ACM International Conference on Automated Software Engineering (ASE), 2020, pp. 1382–1384.
\bibitem{ex6}F. D. Giraldo, B. Milton C., and C. E. Gamboa, ‘Electronic Voting Using Blockchain And Smart Contracts: Proof Of Concept’, IEEE Latin America Transactions, vol. 18, no. 10, pp. 1743–1751, 2020.
\bibitem{ex7}E. M. Sifra, ‘Security Vulnerabilities and Countermeasures of Smart Contracts: A Survey’, in 2022 IEEE International Conference on Blockchain (Blockchain), 2022, pp. 512–515.
\bibitem{ex8}S. S. Kushwaha, S. Joshi, D. Singh, M. Kaur, and H.-N. Lee, ‘Systematic Review of Security Vulnerabilities in Ethereum Blockchain Smart Contract’, IEEE Access, vol. 10, pp. 6605–6621, 2022.
\bibitem{ex9}B. Thuraisingham, ‘Blockchain Technologies and Their Applications in Data Science and Cyber Security’, in 2020 3rd International Conference on Smart BlockChain (SmartBlock), 2020, pp. 1–4.
\bibitem{ex10}M. Maffei, ‘Formal Methods for the Security Analysis of Smart Contracts’, in 2021 Formal Methods in Computer Aided Design (FMCAD), 2021, pp. 1–2.
\bibitem{ex11}S. Sayeed, H. Marco-Gisbert, and T. Caira, ‘Smart Contract: Attacks and Protections’, IEEE Access, vol. 8, pp. 24416–24427, 2020.
\bibitem{ex12}K. B. Kim and J. Lee, ‘Automated Generation of Test Cases for Smart Contract Security Analyzers’, IEEE Access, vol. 8, pp. 209377–209392, 2020.


\end{thebibliography}
\end{document}